\documentstyle[12pt,axodraw]{article}
\newcommand{\ben}{\begin{displaymath}}
\newcommand{\een}{\end{displaymath}}
\newcommand{\be}{\begin{equation}}
\newcommand{\ee}{\end{equation}}
\newcommand{\bea}{\begin{eqnarray}}
\newcommand{\eea}{\end{eqnarray}}

\begin{document}

\begin{center}
{\Large On the Gauge Invariance of the $Z$-Boson Mass\footnote{
The results were obtained at the High Energy Physics Institute of Tbilisi
State University, Tbilisi, Georgia (March 1992).}
}

\vspace{1cm}

{\large  J.~Gegelia$^a$,
G.~Japaridze$^b$,  A.~Tkabladze$^c$, A.~Khelashvili$^d$ \\
and K.~Turashvili$^d$ }

\end{center}

\vspace{1cm}

\begin{center}

{\it $^a$ School of Physical Sciences, Flinders University of South Australia,\\ 
Bedford Park, S.A. 5042, Australia}\\

\vspace{3mm}
{\it $^b$ Center for Theoretical studies of Physical
systems, \\ Clark
Atlanta University, Atlanta, GA 30314, U.S.}\\

\vspace{3mm}
{\it $^c$ DESY Zeuthen, D-15738 Zeuthen, Germany}\\ 

\vspace{4mm}
{\it $^d$ Tbilisi State University, 2 Chavchavadze Ave., \\
Tbilisi 380096, Georgia}
\date{\today}

\end{center}

\vspace{0.5cm}
\begin{abstract}
The different schemes for the definition of the $Z$ boson mass are analyzed.
It is shown that the scheme, defining the mass as  pole of the real
part of the $Z$ boson propagator and the width as the imaginary part of the propagator
at the same point results in the gauge dependent results for these parameters in a two-loop
approximation. On the other hand, the scheme, where the mass and width are related
to the position of the pole of the propagator in the complex plane leads
to the gauge independent result. It is argued that the gauge dependence of mass and width 
does not contradicts to the gauge invariance of the amplitude. 
\end{abstract}
\newpage

\section{Introduction}
Recently the problem of the definition of $Z$ boson mass and width 
has been discussed \cite{willen,stuart}. 
The problem refers to the definition of the physical  
parameters of
unstable particle. The $Z$ boson's {\it in} and {\it out} states can not be considered
as the eigenvectors of the energy-momentum operator (clearly this is true for any
unstable system). Thus we have no unambiguous theoretical prescription for the definition
of $Z$ boson mass. The conventional way is to extract  the mass and the width from the amplitudes of
processes containing the corresponding resonance in the $S$-channel. E.g. the amplitude
of the process $e^{+}e^{-}\rightarrow f\bar{f}$ near its peak position is often parameterized
as \cite{stuart,fuji}:
\begin{equation}
A(S)=\frac{{\cal R}}{S-M_{Z}^{2}+i\Gamma_{Z}M_{Z}}+r(S)
\end{equation}
or
\begin{equation}
A(S)=\frac{{\cal R}}{S-(M_{Z}-i\Gamma_{Z}/2)^{2}}+r(S),
\end{equation}
where ${\cal R}$ is the residue, the remainder, less singular part of the amplitude is
designated by $r(S)$ and it is assumed that the first term numerically exceeds $r$.
The parameterization (1)-(2) may be obtained from the conditions on the Green function
of $Z$ boson field.
These conditions define the renormalization scheme, and as a consequence, the 
resonance parameters.

In consideration of the mass and width two schemes are prevalent.

In the framework of the first scheme the resonance parameters are defined as 
$M_{Z}^{2}=Re\;S_{p}$, $M_{Z}\Gamma_{Z}=Im\;S_{p}$ \cite{stuart}, where $S_{p}$ is 
the position of the pole of propagator in a a complex $S$-plane
\cite{stuart}:
\begin{equation}
D^{-1}(S_{p})=0
\end{equation}
In (2) $D^{-1}$ is the denominator of the exact propagator of the $Z$ boson.

In the another, so called on mass-shell scheme, the variable $S$ remains real and 
the definition is following \cite{fuji}:
\begin{equation}
Re\;D^{-1}(M_{Z}^{2})=0,\;M_{Z}\Gamma_{Z}=Im\;D^{-1}(M_{Z}^{2})
\end{equation}

Of course, many different schemes can be introduced. E.g. the mass can be defined as the peak
position of the amplitude, but in this case it would depend on the process under
consideration. From the theoretical point of view it is preferable  
that the mass of an unstable particle,
in a full analogy with the case of stable one, is defined without referring to any particular 
element of the $S$ matrix but the propagator. 
The condition imposed on the propagator of an unstable particle is a realization 
of the scheme
defining  mass and width.

In this paper we investigate the gauge dependence of the resonance parameters, defined 
in schemes (3) and (4).
Note, that the gauge independence of the physical
mass and width is the separate requirement, not being the direct consequence of the gauge
independence of amplitude (1)-(2). In other words, 
from the requirement of gauge independence
of observables (in our case - $Z$ line shape) it does not 
follow necessarily 
that the resonance parameters should be gauge
invariant. In discussing the gauge dependence one expects that since
the pole of the amplitude is a physical quantity,
the mass and the width, defined through scheme (3) are gauge independent, while it is not
evident in case of scheme (4). Below we check this argument in the framework of perturbation
theory.

In section {\bf 2}  we demonstrate that $M_{Z}$ and $\Gamma_{Z}$, defined in
scheme (3),  are gauge independent in the two-loop approximation
while the scheme (4) leads to the gauge dependence of these parameters
(from \cite{stuart} we have learned that A. Sirlin has arrived to the same conclusion about 
the scheme (4).
Unfortunately at the present time the publication \cite{sirlin} is not available to us  
and hence we are not able 
to compare
our arguments).

In section {\bf 3} it is argued that if we use the scheme (4) or any other scheme with the gauge dependent
$M_{Z}$ and $\Gamma_{Z}$, the scattering amplitude is still gauge independent.

We use the bare Lagrangian in which the Higgs field condensate 
$v$ and the $ZA$ mixing are
chosen in the tree approximation \cite{fuji}. The input parameters are the bare
electromagnetic charge $e_{0}$ and the bare masses $M_{0i}$. Calculations are performed in 
$R_{\xi}$ gauge in the framework of dimensional regularization. We use Feynmann rules from the 
review \cite{fuji}.

\section{Two-loop Analysis of Gauge Dependence}

To investigate the gauge dependence of $M_{Z}$ and $\Gamma_{Z}$ defined in the scheme (3) let
us express $S_{P}$ as follows:
\begin{equation}
S_{P}=M_{0Z}^{2}+e_{0}^{2}\delta M_{1}^{2}+e_{0}^{4}\delta M_{2}^{2}+...,
\end{equation}
where the complex numbers $\delta M_{i}$ do not depend on $e_{0}$. 

It is straightforward to show that the contributions from terms with the gauge parameters 
$\alpha_{W}$,
$\alpha_{Z}$ and $\alpha_{A}$ are factorized in the expression for $\delta M_{1}^{2}$.
The most time-taking part of calculations in
$R_{\xi}$ gauge is
that with the gauge parameter $\alpha_{W}$ \cite{fuji}. Therefore, in the two-loop 
approximation where
there is no factorization we will investigate only the sector with $\alpha_{W}$ to which
we refer throughout
this paper as ${\alpha}$. 

This parameter appears in the free propagators of $W$, $\chi$ and $c$ 
(for the denotions see \cite{fuji}; $\chi$ and $c$ correspond to the
 pseudogoldstone and ghost degrees of freedom appearing in $R_{\xi}$ gauge):
\begin{equation}
D^{\mu\nu}_{W}(p)=\frac{1}{p^{2}-M_{0W}^{2}}\Biggl ( g^{\mu\nu}-
\frac{p^{\mu}p^{\nu}}{M_{0W}^{2}} \Biggr)
+\frac{p^{\mu}p^{\nu}}{M_{0W}^{2}(p^{2}-\alpha M_{0W}^{2})}
\end{equation}
and
\begin{equation}
D_{\chi}(p)=-\frac{1}{p^{2}-\alpha M_{0W}^{2}},\;D_{c^{\pm}}(p)=
-\frac{1}{p^{2}-\alpha M_{0W}^{2}}
\end{equation} 
For brevity we will term the denominators with $\alpha$ as $\alpha$-terms.

One loop correction to the $Z$ boson self energy contains terms, 
linear in $\alpha$ (i.e. integrand contains only one $\alpha$-denominator)
 as well as $\alpha^{2}$-terms (the product of two $\alpha$-denominators).
 Two loop correction contains $\alpha$, $\alpha^{2}$ 
and $\alpha^{k}$, $k>2$ terms. The simple calculation based on Ward
 identities demonstrates that $\alpha^{k}$, $k>2$ terms cancel trivially, 
and the technique is exactly the same as in the 
case of stable particles. Further we will discuss only $\alpha$ and
$\alpha^{2}$-terms which are present in one-loop and two-loop
 corrections and interfere with each other.\\ 
%
\vspace{0.5cm}
\begin {figure} [htbp]
\begin{picture}(400,100)
\Line(30,50)(50,50)
\Line(110,50)(130,50)
\BCirc(80,50){30}
\BCirc(220,50){30}
\Line(190,20)(250,20)
\Line(320,20)(360,20) 
\Line(340,20)(340,40)
\BCirc(340,60){20}
\Text(30,40)[l]{$ Z$}
\Text(130,40)[r]{$ Z$}
\Text(90,85)[l]{$ W,\chi,c$}
\Text(80,0)[l]{a)}
\Text(180,20)[l]{$Z$}
\Text(255,20)[l]{$Z$}
\Text(230,85)[l]{$ W,\chi,c$}
\Text(220,0)[l]{b)}
\Text(310,20)[l]{$Z$}
\Text(365,20)[l]{$Z$}
\Text(350,85)[l]{$ W,\chi,c$}
\Text(340,0)[l]{c)}
\Text(345,33)[l]{$\phi$}
\end{picture}
\vspace*{2mm}\\\vspace*{0.0cm}\\
\caption{One loop contribution from the "$\alpha$ - sector" in $Z$ boson self energy.}
\end{figure}
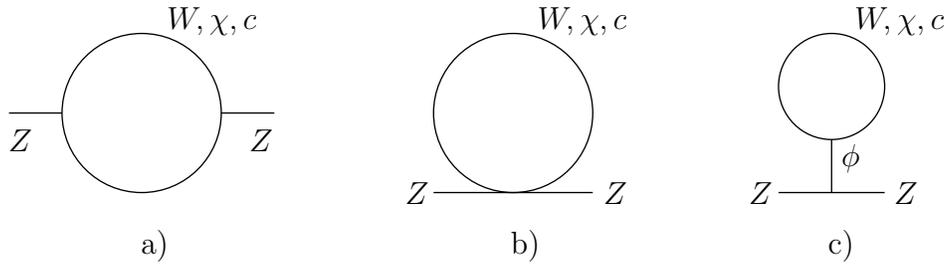
%
The $\alpha^{2}$-terms appear only in the diagram Fig.1,a) and all the diagrams contain terms,
linear in $\alpha$. Even now two-loop calculations are still tedious.

To proceed further let us note that in the sense of the gauge dependence the crucial
difference between the results following from the schemes (3) and (4) is caused by fermion 
loops, leading to $Im\;\delta M_{1}^{2}\neq 0$. Since our goal is to establish the (possible) 
gauge dependence of $M_{Z}$ and $\Gamma_{Z}$, in the gauge independent part of 
$\delta M_{1}^{2}$ we retain the contribution from the lepton sector only. The 
quark loop contribution gives the similar result.

Summarizing, we consider only $\alpha$ and $\alpha^{2}$ terms from the 
"$\alpha$-sector" and 
in the expression for $\delta M_{1}^{2}$ only the contributions from the electron and neutrino 
loops are taken into account.

After these introductory remarks we are in a position to discuss the gauge dependence of 
$M_{Z}$ and $\Gamma_{Z}$. 

Let us 
consider first the one-loop contribution to $\Pi_{\mu\rho}$ - self energy 
of the $Z$ boson. To extract parameters $M_{Z}$ and $\Gamma_{Z}$ it is enough to consider 
the coefficient of $g_{\mu\rho}$. Therefore we will follow only the integrals, leading 
to this tensor structure. The corresponding contribution of $\alpha$ and $\alpha^{2}$ 
terms in $\Pi_{\mu\rho}$ is: 
\begin{eqnarray}
\Pi_{\mu\rho}(p, \alpha)=-e_{0}^{2}\frac{(p^{2}-M_{0Z}^{2})(p^{2}+M_{0Z}^{2})}
{M_{0W}^{2}(M_{0Z}^{2}-M_{0W}^{2})}J_{1\mu\rho}-
\cr e_{0}^{2}g_{\mu\rho}\biggl[ 2\frac{M_{0Z}^{2}-p^{2}}
{M_{0Z}^{2}-M_{0W}^{2}}J_{1}+4(M_{0Z}^{2}-p^{2})J_{2} \biggr]-
 e_{0}^{2}\frac{4(M_{0Z}^{2}-p^{2})}{M_{0W}^{2}}J_{2\mu\rho},
\end{eqnarray}
where $p$ is the $Z$ boson momentum  and
\begin{equation}
J_{1\mu\rho}=i\int d^{n}q\; \frac{q_{\mu}q_{\rho}}{(q^{2}-\alpha M_{0W}^{2})((p+q)^{2}-
\alpha M_{0W}^{2})},
\end{equation}
\begin{equation}
J_{2\mu\rho}=i\int d^{n}q\; \frac{q_{\mu}q_{\rho}}{(q^{2}-M_{0W}^{2})((p+q)^{2}-
\alpha M_{0W}^{2})},
\end{equation}
\begin{equation}
J_{1}=i\int d^{n}\; \frac{1}{q^{2}-\alpha M_{0W}^{2}},
\end{equation}
\begin{equation}
J_{2}=i\int d^{n}q\; \frac{1}{(q^{2}-M_{0W}^{2})((p+q)^{2}-\alpha M_{0W}^{2})}
\end{equation}
In (9)-(12) $n$ is a space-time dimension. In the one-loop approximation 
$\Pi_{\mu\rho}$ can be 
expressed as
\begin{equation}
\Pi_{\mu\rho}=g_{\mu\rho}(p^{2}-M_{0Z}^{2})-e_{0}^{2} \biggl[ \Pi_{\mu\rho}^{(1)inv}(p^{2})+
\Pi_{\mu\rho}(\alpha) \biggr],
\end{equation}
where $\Pi_{\mu\rho}^{(1)inv}=g_{\mu\rho}\Pi^{(1)inv}$ and $\Pi_{\mu\rho}(\alpha)$ are
correspondingly
gauge independent and gauge dependent parts. The gauge independence of $M_{Z}$ and $\Gamma_{Z}$
in one-loop approximation is evident in both schemes, since in the scheme (3), where 
$p^{2}=S_{p}$ we have 
\begin{equation}
S_{P}-M_{0Z}^{2}=e_{0}^{2}\Pi^{(1)inv}(M_{0Z}^{2}),
\end{equation}
and in the scheme (4) where $p^{2}=M_{Z}^{2}$, we obtain:
\begin{eqnarray}
M_{Z}^{2}-M_{0Z}^{2}=e_{0}^{2}Re\;\Pi^{(1)inv}(M_{0Z}^{2}),\cr
M_{Z}\Gamma_{Z}=e_{0}^{2}Im\;\Pi^{(1)inv}(M_{0Z}^{2})
\end{eqnarray}

Let us  proceed to the order $e_{0}^{4}$. The self energy in the scheme (3) can be  expressed
as
\begin{equation}
\Pi_{\mu\rho}(S_{P},\alpha)=-e_{0}^{4}\Biggl( \Pi^{(1)inv}(M_{0Z}^{2}) \Biggr)^{2}I_{\mu\rho}+
O(e_{0}^{6}),
\end{equation}
where
\begin{eqnarray}
I_{\mu\rho}=\frac{2M_{0Z}^{2}}{M_{0W}^{2}(M_{0Z}^{2}-M_{0W}^{2})}\;J_{1\mu\rho}\;-\cr
g_{\mu\rho}\Biggl( \frac{2}{M_{0Z}^{2}-M_{0W}^{2}}J_{1}+4J_{2} \Biggr) - \frac{4}{M_{0W}^{2}}\;
J_{2\mu\rho}
\end{eqnarray}
and in the scheme (4) $Re\;\Pi_{\mu\rho}$ can be expressed as
\begin{equation}
Re\;\Pi_{\mu\rho}(\alpha,p^{2})|_{p^{2}=M_{Z}^{2}}=-e_{0}^{4}Re\;\Pi^{(1)inv}Re\;I_{\mu\rho}
\end{equation}
Let us remind that in $\Pi_{\mu\rho}^{(1)inv}$ we consider contributions
only from the electron and neutrino.
Then, to investigate the gauge dependence in order $e_{0}^{4}$ it is necessary to consider
the two-loop  diagrams, Fig. 2 a) and b):\\
%
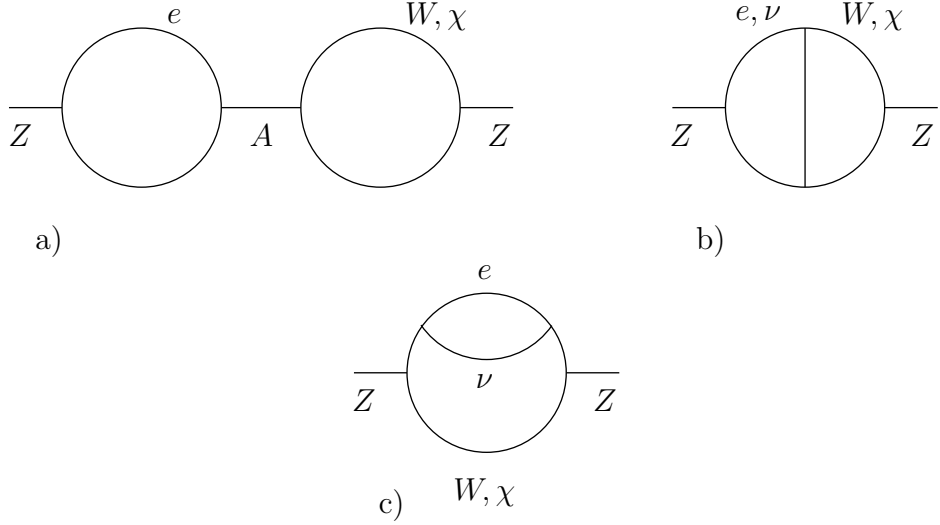
\begin {figure} [t]
\begin{picture}(400,200)
\Line(30,150)(50,150)
\Line(110,150)(140,150)
\BCirc(80,150){30}
\BCirc(170,150){30}
\Line(200,150)(220,150)
\Line(280,150)(300,150) 
\Line(360,150)(380,150)
\BCirc(330,150){30}
\Line(330,120)(330,180)
\Line(160,50)(180,50) 
\Line(240,50)(260,50)
\BCirc(210,50){30}
\CArc(210,85)(30,-145.5,-35)
\Text(30,140)[l]{$ Z$}
\Text(130,140)[r]{$ A$}
\Text(220,140)[r]{$ Z$}
\Text(90,185)[l]{$ e$}
\Text(180,185)[l]{$ W,\chi$}
\Text(40,100)[l]{a)}
\Text(280,140)[l]{$Z$}
\Text(380,140)[r]{$Z$}
\Text(305,185)[l]{$ e,\nu$}
\Text(345,185)[l]{$W,\chi$}
\Text(290,100)[l]{b)}
\Text(160,40)[l]{$Z$}
\Text(260,40)[r]{$Z$}
\Text(210,5)[c]{$W,\chi$}
\Text(210,88)[c]{$e$}
\Text(170,0)[l]{c)}
\Text(210,47)[c]{$\nu$}

\end{picture}
\vspace*{2mm}\\
\caption{Two-loop contribution from "$\alpha$ - sector" in $Z$ boson self energy.}
\end{figure}
The contribution from $\alpha^{2}$-terms in $\Pi_{\mu\rho}$ from the diagram Fig. 2 a) at
$p^{2}=M_{0Z}^{2}$ is equal to:
\begin{equation}
-ie_{0}^{4}\frac{M_{0Z}^{4}}{4M_{0W}^{4}(M_{0Z}^{2}-M_{0W}^{2})}J_{1\mu\sigma}
\int d^{n}q\; tr\biggl[ G(p+q)\Gamma_{\sigma}G(q)\gamma_{\rho} \biggr],
\end{equation}
where $G(p)=1/(m_{0}-\hat{p})$ is the free electron propagator and
\begin{equation}
\Gamma_{\sigma}=\gamma_{\sigma}\biggl[ -1+4\frac{M_{0Z}^{2}-M_{0W}^{2}}{M_{0Z}^{2}}+
\gamma_{5} \biggr]
\end{equation}
is the $Ze^{+}e^{-}$ vertex.
Let us discuss the cancelation technique for the terms similar to (19) for the diagram
Fig. 2 b), containing two $W$ bosons. First of all,  we  consider the fermion loop contribution
multiplied on virtual $W$ boson's momenta that correspond to $\alpha^{2}$-terms:
\begin{eqnarray}
\int \frac{d^{n}kd^{n}q}{(k^{2}-\alpha M_{0W}^{2})((p+k)^{2}-\alpha M_{0W}^{2})}
\Gamma_{\sigma\beta\rho}(-k,p+k,-p)\cr
tr \biggl[ G(p+q)\Gamma_{\mu}G(q)\hat{k}(1-\gamma_{5})S(q-k)(\hat{p}+\hat{k}) \biggr],
\end{eqnarray}
where $\Gamma_{\sigma\beta\rho}$ is the $ZW^{+}W^{-}$ vertex and $S(p)=-1/\hat{p}$ is the
free neutrino propagator.
Using $\hat{k}=m_{0}-G^{-1}(q)+S^{-1}(q-k)$ and $\hat{p}+\hat{k}=m_{0}-
G^{-1}(p+q)+S^{-1}(q-k)$ for the trace in (19) we obtain:
\begin{eqnarray}
tr \biggl[ G(p+q)\Gamma_{\mu}G(q)(1+\gamma_{5})(\hat{p}+\hat{k})+
m_{0}^{2}G(p+q)\Gamma_{\mu}G(q)(1-\gamma_{5})S(q-k)+\cr
m_{0}G(p+q)\Gamma_{\mu}G(q)(1-\gamma_{5})-
m_{0}G(p+q)\Gamma_{\mu}(1-\gamma_{5})S(q-k)-\cr
m_{0}\Gamma_{\mu}G(q)(1-\gamma_{5})S(q-k)  \biggr]
\end{eqnarray}
The terms with the three fermion propagators are canceled at $p^{2}=M_{0Z}^{2}$ in the sum
of all the diagrams of type Fig. 2 b). Without this cancelation the gauge dependence of  
$M_{z}$ and $\Gamma_{Z}$ is unavoidable.
The terms in (22), containing the product of propagators $G$ and $S$ are
canceled by the $\alpha^{2}$-terms, originated from the diagram Fig. 2 c).
Now, the sum of diagrams Fig.2 a) and b) is
%
\begin{eqnarray}
ie_{0}^{4}\;\frac{M_{0Z}^{6}J_{1\rho\sigma}}{16M_{0W}^{2}(M_{0Z}^{2}-M_{0W}^{2})}\int d^{n}q\;
tr \biggl[ G(p+q)\Gamma_{\mu}G(q)\Gamma_{\sigma}+\cr
S(p+q)\gamma_{\mu}(1-\gamma_{5})S(q)\gamma_{\rho}(1-\gamma_{5})  \biggr]
\end{eqnarray}
%
It is straightforward to demonstrate that the terms in (23), leading to the tensor structure
$g_{\mu\rho}$ can be expressed as:
%
\begin{eqnarray}
e_{0}^{4}\;\frac{2M_{0Z}^{2}}{ M_{0W}^{2}(M_{0Z}^{2}-M_{0W}^{2})}\biggl[
\Pi_{(e)}^{(1)inv}(M_{0Z}^{2})+\Pi_{(\nu)}^{(1)inv}(M_{0Z}^{2}) \biggr]J_{1\mu\rho},
\end{eqnarray}
%
where $\Pi_{(e)}^{(1)inv}$ and $\Pi_{(\nu)}^{(1)inv}$ refer to contributions of electron 
and neutrino.
 The $\alpha$-terms is canceled in the sum of the expressions (16) and (24), and it turns out that $S_p$
is gauge invariant in order of  $e_{0}^{4}$. On the other hand, the sum of the real parts of (18) and (23)
still depends on $\alpha$.
Therefore, the scheme (3), in distinct with (4), defines the gauge independent parameters
$M_{Z}$ and $\Gamma_{Z}$.

We have checked the gauge independence of $S_{p}$ also in the framework of conventional covariant
gauge, when the gauge fixing term is ${\partial_{\mu}W_{\mu}}^{2}/2\alpha$. In this case
the diagrams type of Fig. 1 b) and c) are gauge independent.
The cancelation of the gauge dependent terms occurs in a full analogy
with the case of $R_{\xi}$ gauge.

A few words about $ZA$ mixing: when we define the mixing scheme by the requirement that Green's
function $\langle ZA\rangle$ has no poles (i.e. the mass-shell for the photon is defined
only from $\langle AA \rangle$ and $S_{p}$  is defined only from $\langle ZZ \rangle$), there
arise a new effective vertices but $S_{p}$ does not change. Note that the same value
of this parameter
is obtained as a position of pole of the Green's functions $\langle aa \rangle$,
$\langle B_{3}B_{3}  \rangle$ and $\langle aB_{3}  \rangle$, where $a$ and $B_{3}$ are the
original gauge fields of the gauge groups $U(1)$ and $SU(2)$ leading after mixing to
$A$ and $Z$.

Note also that if  we take into account the quantum corrections in the
definition of the Higgs condensate in the bare Lagrangian,
 i.e. $v=v_{0}+\delta v$, then $\langle \phi \rangle=0$,
but the parameter $S_{p}$ would be the same. Moreover, the replacement $v_{0}\rightarrow v$
does not affect the Green's function which does not  contain
the Higgs field $\phi$ on external lines.

\section{Gauge dependence of $M_{Z}$ and $\Gamma_{Z}$ and the Gauge Invariance 
of the Amplitude}

As we have demonstrated in section {\bf 2},  the parameters $M_{Z}$ and
 $\Gamma_{Z}$ defined in scheme (3)
do not depend on the gauge parameter in the two-loop approximation. Since the
parameters $M_{Z}$ and $\Gamma_{Z}$ defined in the scheme (4) are gauge dependent, it might seem
that the scheme (3) is preferable. 
Below we show that the requirement of the gauge independence as well as the requirement of
numerical convergence does not lead in principle to the preference of one 
particular scheme.

Let us express $M_{Z}$ and $\Gamma_{Z}$ in the framework of regularized theory (i.e. $n\neq 4$)
as follows:
%
\begin{equation}
M_{Z}=M_{0Z}\sum^{\infty}_{k=1}e_{0}^{2k}\delta_{1k},\;\Gamma_{Z}=e_{0}^{2}M_{Z}^{n-3}
\sum^{\infty}_{k=1}e_{0}^{2k}\delta_{2k}
\end{equation}
%
It is possible to resolve $M_{0Z}$ and $e_{0}^{2}$ in terms of $M_{Z}$ and $\Gamma_{Z}$:
%
\begin{equation}
M_{0Z}=M_{Z}\sum^{\infty}_{k=1} \Biggl( \frac{\Gamma_{Z}}{M_{Z}} \Biggr)^{k}\triangle_{1k}=M_{Z}
Z_{M_{Z}},
\end{equation}
%
and
%
\begin{equation}
e_{0}^{2}=\frac{\Gamma_{Z}}{M_{Z}^{n-3}}\sum^{\infty}_{k=1} \Biggl( \frac{\Gamma_{Z}}{M_{Z}} 
\Biggr)^{k}
\triangle_{2k}=\Gamma_{Z}Z_{\Gamma_{Z}},
\end{equation}
%
where $Z_{M_{Z}}$ and $Z_{\Gamma_{Z}}$ are  the mass and width renormalization factors.

Clearly, the coefficients in expressions (25) - (27) are gauge independent when calculated
in the scheme (3) and depend on a gauge in the case of  scheme (4).

If we substitute (26)-(27) in the expression for the physical quantity, e.g. the amplitude 
$\langle e^{+}e^{-}|f\bar{f} \rangle$ we obtain
the ultraviolet finite result which may be parameterized as (1)-(2), or as follows:
%
\begin{equation}
A(S)=\sum^{\infty}_{k=1} \Biggl( \frac{\Gamma_{Z}}{M_{Z}} \Biggr)^{k}a_{k}(S)
\end{equation}
%

The gauge independence of $A(S)$ in a scheme (3) is evident since the coefficients $a_{k}(S)$
do not depend on gauge parameters. In  scheme (4) the coefficients  $a_{k}(S)$ contain $\alpha$
 explicitly but this does not lead to the gauge dependence of amplitude $A(S)$. Indeed,
for this case, in the expression
%
\begin{equation}
\frac{dA}{d\alpha}=\frac{\partial A}{\partial \alpha}+
\frac{\partial A}{\partial \Gamma_{Z}}\frac{d\Gamma_{Z}}{d\alpha}+
\frac{\partial A}{\partial M_{Z}}\frac{dM_{Z}}{d\alpha}
\end{equation}
%
every term is non vanishing but in the sum they cancel since the scheme (4) is connected with
the scheme (3) via the renormalization group equations with respect to the parameter $\alpha$.
These equations can be obtained from (26)-(27):
%
\begin{equation}
\frac{dM_{Z}}{d\alpha}=\beta_{M_{Z}}M_{Z},\;\frac{d\Gamma_{Z}}{d\alpha}=
\beta_{\Gamma_{Z}}\Gamma_{Z},
\end{equation}
%
where
%
\begin{equation}
Z_{M_{Z}}\beta_{M_{Z}}\equiv -\frac{dZ_{M_{Z}}}{d\alpha},\;Z_{\Gamma_{Z}}\beta_{\Gamma_{Z}}
\equiv -\frac{dZ_{\Gamma_{Z}}}{d\alpha}
\end{equation}
%
The similar equations for the other masses $M_{i}$ can be easily obtained. It is evident that
if we consider the scheme where $\beta_{M,\Gamma}\neq 0$, the independence of physical quantities
on $\alpha$ appears in a full analogy with the independence on a renormalization point.
If we use (29) and (30),  we would obtain from (28) that
%
\begin{equation}
\frac{d}{d\alpha}\sum^{N}_{k=1}\Biggl ( \frac{\Gamma_{Z}}{M_{Z}} \Biggr)^{k}a_{k}(S)=
O \biggl[ \Biggl (\frac{\Gamma_{Z}}{M_{Z}} \Biggr)^{N+1} \biggr]
\end{equation}
%
in each order of $\Gamma_{Z}$.
Clearly, the same relations hold for the remaining gauge parameters. Relation similar to
(32) with respect to renormalization point $\mu$ states that (see, e.g. \cite{bjorken}):
%
\begin{equation}
\frac{dP_{N}(\mu,\;g(\mu))}{d\mu}\equiv \frac{d}{d\mu}\sum^{N}_{k=1} g^{k}(\mu)P_{k}=
O \Biggl (g(\mu)^{N+1} \Biggr),
\end{equation}  
%
where $P_{N}$ is a physical quantity calculated up to order $N$-th order in renormalized 
coupling $g(\mu)$. In any particular 
order of perturbation theory the response of a physical quantity on $\mu\rightarrow \mu+d\mu$ 
is of a {\it next order} in $g(\mu)$. Evidently, the exact expression is the 
renormalization point (and gauge parameter) independent. 
So, the requirement of the gauge independence of the amplitude
$\langle e^{+}e^{-}|f\bar{f} \rangle$  is not sufficient to prefer scheme (3).

Let us consider the problem  of numerical analysis. The peak position
of the amplitude, $S_{0}$, is defined as the solution of the equation:
%
\begin{equation}
\frac{d}{dS}\mid A(S)\mid=0
\end{equation}
%
It is easy to show that in a scheme (3) we obtain
%
\begin{equation}
S_{0}=M_{Z}^{2}+\frac{\Gamma_{Z}^{3}}{M_{Z}^{3}}\sum^{\infty}_{k=0}
\Biggl ( \frac{\Gamma_{Z}}{M_{Z}} \Biggr)^{k}C_{k}(M_{i}),
\end{equation}
%
while the scheme (4) leads to
%
\begin{equation}
S_{0}=M_{Z}^{2}+\frac{\Gamma_{Z}^{2}}{M_{Z}^{2}}\sum^{\infty}_{k=0}
\Biggl ( \frac{\Gamma_{Z}}{M_{Z}} \Biggr)^{k}C^{\star}_{k}(M_{i})
\end{equation}
%
Evidently, if the series (35) and (36) converge fast enough, the scheme (3) is preferable when
$\Gamma_{Z}\ll M_{Z}$. But from the theoretical point of view this  is rather troublesome to 
confirm. Indeed, the coefficient functions $C_{k}$ and $C^{\star}_{k}$ depend on the masses of  all
 the
particles of the standard model, so their values can not be estimated with the help of the
current data.

Note that there  exists other scheme which leads to the same results in a two-loop 
approximation as the scheme (3).
For example one can define a new scheme by the relation:
\begin{equation}
\frac{d}{dS}\mid D^{-1}(S) \mid_{S=M_{Z}^{2}}=0
\end{equation}
The simple calculation shows that $M_{Z}$, defined by (37) differs from the result of scheme
(3) only in order $e_{0}^{6}$. In other words, in a two-loop approximation the scheme (37)
leads to the gauge independent $M_{Z}$ and this $M_{Z}$ satisfies relation, similar to
(3). The results following from these two schemes differ only in three-loop approximation.

\section{Discussion}

In this paper, based on the gauge independence and numerical analysis we have investigated the
schemes defining the mass and the width of $Z$ boson.

The relations (35) and (36) demonstrate that the schemes (3) and (37) may be preferable. On the
other
hand, since in a scheme (4) the arbitrary parameter $\alpha$ remains, the problem of
convergence in (36) requires the consideration of the renormalization group equations (30). It
may occur that for some numerical values of $\alpha$ the series (36) would converge faster
than the series (35). It is not surprising since from (29)-(30) it is evident that the
gauge parameter
${\alpha}$
plays the role, analogous of renormalization point and thus, in principle, the variation of 
${\alpha}$ can improve the convergence of series in $\Gamma_{Z}/M_{Z}$. 
Hence, the requirement of the numerical convergence
can not
be used in defining the physical mass and the width of the $Z$ boson, since this requirement
may be in contradiction with requirement of gauge invariance of these parameters. Note also that
the result of numerical analysis depends essentially on the process under
consideration.

 From the requirement of gauge independence the schemes (3) and (37) are preferable. Unfortunately,
this principle is necessary but not sufficient and therefore does not leads to the unique choice
of a scheme. Indeed, besides (3) and (37), infinitely many schemes for defining the gauge
independent parameters can be pointed out. For example let us express the $Z$ boson propagator
as
\begin{equation}
D=D^{inv}+D^{\star},
\end{equation}
where $D^{inv}$ stands for the contribution of physical degrees of freedom (i.e. $D^{inv}$ is
gauge independent) and $D^{\star}$ varies from gauge to gauge. Consider now any conditions on
real and imaginary parts of $D^{inv}$, leading to the ultraviolet renormalization. It is
evident that these conditions define two gauge invariant parameters which may be treated as
$M_{Z}$ and $\Gamma_{Z}$.

Despite of this non uniqueness, the schemes (3) and (37) are remarkable.
Indeed, $M_{Z}$ and $\Gamma_{Z}$ defined
in scheme (3) are connected with the position of the pole of propagator on the complex plane and
in scheme (37) they can be obtained from the position of the peak of amplitude on real axis.
Apparently it is the scheme (3) that defines the physical mass and the width of
$Z$ boson, since, from our point of view, only this scheme allows us to formulate the
Lehmann-Symanzik-Zimmermann (LSZ) reduction technique \cite{bjorken} for the processes 
involving $Z$ boson in
initial or final states.

Some papers contain attempts to generalize LSZ technique for unstable particles 
(see e.g \cite{hammer}) but
we think that more thoroughful analysis is necessary. 
Our recent calculations show that the residue of the amplitude
$\langle e^{+}e^{-}|f\bar{f} \rangle$  in a two-loop approximation is gauge independent
at $S=S_{p}$. This fact already indicates the incontrovertible preference of scheme (3)
since the gauge independence of the residue apparently leads to the gauge independent
amplitude for the decay $Z\rightarrow f \bar{f}$.



\vspace*{1cm}
These results were presented at the Conference "Quarks-92", helded in 
Zvenigorod  Russia, 1992.
Since then the accuracy in defining of the $Z$-boson line shape has been
 substantially improved what require the consideration of the higher order 
corrections. Therefore, our results are still actual and worth appearing in 
the Archive. The text of original version is slightly changed.\\
The preprint [4] is published in:\\
A.Sirlin, {\it Phys. Rev. Lett} {\bf 67}, 2127, (1991).

\end{document}